\documentclass[11pt,a4paper]{article}
\usepackage{amsmath}
\usepackage{epsfig}
\author{H. Mohseni Sadjadi$^1$\footnote{mohseni@phymail.ut.ac.ir} and M.
Honardoost$^2$
\\$^1$ {\small School of Physics, University College of Science,  University of Tehran,}
\\ {\small North Karegar Ave.  Tehran, Iran.}
\\$^2$ {\small Department of Physics, Shahid Beheshti University,}
\\ {\small Evin, Tehran 19839, Iran.}}

\title{Thermodynamics second law and $\omega=-1$ crossing(s) in interacting
holographic dark energy model }
\begin{document}
\maketitle
\begin{abstract}
By the assumption that the thermodynamics second law is valid, we
study the possibility of $\omega=-1$ crossing in an interacting
holographic dark energy model. Depending on the choice of the
horizon and the interaction, the transition from quintessence to
phantom regime and subsequently from phantom to quintessence phase
may be possible. The second transition avoids the big rip
singularity. We compute the dark energy density at transition time
and show that by choosing appropriate parameters we can alleviate
the coincidence problem.

\end{abstract}
\section{Introduction}
Recent observations suggest that the universe is undergoing an
accelerated expansion \cite{acc}. This acceleration may be
explained by the assumption that $70\%$ of the universe is filled
by a perfect fluid with negative pressure, dubbed dark energy.
Some present data seem to favor an evolving dark energy,
corresponding to an equation of state (EOS) parameter less than
$\omega=-1$ at present epoch from $\omega>-1$ in the near past
\cite{Bo}. Many candidates for dark energy has been proposed such
as the cosmological constant \cite{dark}: A constant quantum
vacuum energy density which fills the space homogeneously,
corresponding to a fluid with a constant EOS parameter
$\omega=-1$; dynamical fields with a suitably chosen potential to
make the vacuum energy vary with time \cite{field}, and so on.
Recently, using holographic principle, a new candidate for dark
energy which is independent of any specific field has been
suggested \cite{Li},\cite{hol}. Based on the holographic principle
(which relates the number of degrees of freedom of a physical
system to the area of its boundary), in order to allow the
formation of black holes in local quantum field theory, Cohen et
al \cite{cohen} proposed a relationship between UV and IR cutoff.
This yields an upper bound on the zero-point energy density, which
by a suitable choice of the infrared cutoff, can be viewed as the
holographic dark energy density. In \cite{Li}, three candidates
for the infrared cutoff was proposed: the Hubble radius, the
particle horizon and the future event horizon. There was shown
that among these options only the future event horizon may be
identified with the desired infrared cutoff. To study the
coincidence problem and also to have other choices for the
infrared cutoff, e.g. the Hubble radius, interaction between dark
matter and dark energy \cite{inter} may be considered in the
holographic dark energy model. As we have mentioned, based on
astrophysical data, we may take into account the possibility of
$\omega=-1$ (phantom divide line) crossing. Therefore dark energy
models which can describe phantom divide line crossing, has been
also studied vastly in the literature \cite{divide}. The phantom
like behavior of interacting holographic dark energy was studied
in \cite{abd}, where it was claimed that by selecting appropriate
interaction parameters the transition from the dark energy EOS
parameter $\omega_D>-1$ to $\omega_D<-1$ is possible. Despite
this, in \cite{antabd} it was shown that the dark energy effective
EOS parameter cannot cross $\omega_D^{eff.}=-1$.

In this paper we consider interacting holographic dark energy
model and study the ability of the model to describe the
transition from quintessence to phantom regime and vice versa.
After preliminaries in section two, where we introduce the
interacting  holographic dark energy model and some of its general
properties used in the subsequent sections, in section three we
study the possibility of crossing $\omega=-1$. In section four we
derive necessary conditions for existence of two transitions in
our model. The first transition is from quintessence to phantom
phase and the second the transition from phantom to quintessence
regime. The importance of the second transition lies on the fact
that it avoids the big rip singularity. We discuss also the
behavior of Hubble parameter and dark energy density at
transitions times.

We use $\hbar=G=c=k_B=1$ units throughout the paper.
\section{preliminaries}
We consider a spatially flat
Friedmann--Lemaitre--Robertson--Walker (FRW) universe), with scale
factor $a(t)$
\begin{equation}\label{1}
ds^2=dt^2-a^2(t)(dx^2+dy^2+dz^2).
\end{equation}
We assume that this universe is filled with dark energy and
pressureless dark matter fluids satisfying the following equations
of state
\begin{eqnarray}\label{2}
\dot{\rho_D}+3H(1+\omega_D)\rho_D&=&-Q  \nonumber \\
\dot{\rho_m}+3H\rho_m&=&Q.
\end{eqnarray}
$H=\dot{a}/a$ is the Hubble parameter, $\rho_D$ is dark energy
density and $\rho_m$ is the density of cold dark matter. "dot"
denotes derivative with respect to the comoving time. $\omega_D$
is dark energy EOS parameter. $Q$ denotes the interaction of dark
matter with dark energy. In this paper $Q$ is assumed to be
\begin{equation}\label{3}
Q=(\lambda_m\rho_m+\lambda_D\rho_D)H,
\end{equation}
where $\lambda_m$ and $\lambda_D$ are two real constants.
Different choices such as $\lambda_m=0$, $\lambda_D=0$ and
$\lambda_D=\lambda_m$ has been adopted in literature \cite{inter},
\cite{abd}. The Hubble parameter satisfies
\begin{eqnarray}\label{4}
H^2&=&\frac{8\pi}{3}(\rho_m+\rho_D) \nonumber \\
&=&\frac{8\pi}{3}\rho,
\end{eqnarray}
and
\begin{eqnarray}\label{5}
\dot{H}&=&-4\pi\left((1+\omega_D)\rho_D+\rho_m\right) \nonumber \\
&=&-4\pi(1+\omega)\rho.
\end{eqnarray}
$\rho=\rho_m+\rho_D$ is the total energy density satisfying
\begin{equation}\label{6}
\dot{\rho}+3H(1+\omega)\rho=0,
\end{equation}
and $\omega$ is the parameter of the EOS of the universe. In terms
of $\Omega_D=\rho_D/\rho$, we can write
\begin{equation}\label{7}
\omega_D\Omega_D=\omega.
\end{equation}
$\rho_D$ can be viewed as a holographic energy density
\begin{equation}\label{8}
\rho_D=\frac{3}{8\pi}\frac{c^2}{L^2}.
\end{equation}
The length scale $L$ is an infrared cutoff and $c>0$ is a positive
numerical constant. We assume
\begin{equation}\label{9}
L=\beta R_{FH}+\alpha R_{PH},
\end{equation}
where $R_{FH}$ and $R_{PH}$ are the future and particle event
horizons
\begin{eqnarray}\label{10}
R_{FH}&=&a\int_t^\infty\frac{dt}{a}\nonumber \\
R_{PH}&=&a\int_0^t\frac{dt}{a},
\end{eqnarray}
and $\alpha$ and $\beta$ are two positive numerical constants. By
taking $\beta=0$ and $\alpha=1$, we arrive at the holographic
cosmology horizon adopted in \cite{sus}. For $\alpha=0$ and
$\beta=1$, the infrared cutoff becomes the future event horizon
\cite{Li}. Generally one can assume that $L$ is a function of
$R_{FH}$ and $R_{PH}$ \cite{odin}. The time derivative of $L$ is
obtained as
\begin{equation}\label{11}
\dot{L}=HL+\alpha-\beta.
\end{equation}
Using (\ref{4}) and (\ref{8}), $HL$ in the above equation can be
written as
\begin{equation}\label{12}
HL=c\Omega_D^{-\frac{1}{2}}.
\end{equation}
By relating the entropy of the universe to the infrared cutoff $L$
via
\begin{equation}\label{13}
S=\pi L^2,
\end{equation}
the second law of thermodynamics results in $\dot{L}>0$, leading
to $\dot{\rho_D}<0$. In a phantom dominated universe (identified
by $\dot{H}>0$), using (\ref{4}) and (\ref{8}) we get
\begin{equation}\label{14}
\dot{\Omega_D}=\frac{8\pi}{3H^4}\left(\dot{\rho_D}H^2-2H\dot{H}\rho_D\right)<0.
\end{equation}

For $\dot{\Omega_D}>0$ we must have $(LH\dot{)}<0$ which leads to
$\ddot{L}<0$. But if one requires $\dot{L}\geq 0$, then either
$\lim_{t\to \infty}\dot{L}=0$ or $\ddot{L}$ becomes positive after
a finite time.

In terms of the Hubble parameter, $\omega$ is
\begin{equation}\label{15}
\omega=-1-\frac{2\dot{H}}{3H^2}.
\end{equation}
By substituting $\dot{H}=-H^2+(\beta-\alpha)H/L+(HL\dot{)}/L$
(which can be verified using (\ref{11})), (\ref{15}) becomes
\begin{equation}\label{16}
\omega=-\frac{1}{3}-\frac{2}{3}(\beta-\alpha)\frac{\sqrt{\Omega_D}}{c}+\frac{\dot{\Omega_D}}
{3H\Omega_D}.
\end{equation}
Using (\ref{2}) one can show that
\begin{equation}\label{17}
\dot{r}=3rH\left[\omega_D
+\frac{1}{3}\left(\frac{r+1}{r}\right)\left(\lambda_D+r\lambda_m\right)\right],
\end{equation}
where we have defined $r=\rho_m/\rho_D$. Therefore, by considering
$r=\Omega_D^{-1}-1$, we obtain
\begin{equation}\label{18}
\omega_D=-\frac{1}{3H}\frac{\dot{\Omega_D}}{\Omega_D(1-\Omega_D)}-\frac{\lambda_D}{3(1-\Omega_D)}
-\frac{\lambda_m}{3\Omega_D},
\end{equation}
and consequently
\begin{equation}\label{19}
\omega=-\frac{1}{3H}\frac{\dot{\Omega_D}}{(1-\Omega_D)}-\frac{\lambda_D\Omega_D}{3(1-\Omega_D)}
-\frac{\lambda_m}{3}.
\end{equation}
From (\ref{16}) and (\ref{19}) we can express $\omega$ and
$\dot{\Omega_D}$ in terms of $\Omega_D$:
\begin{equation}\label{20}
\omega=-\frac{1+\lambda_D-\lambda_m}{3}\Omega_D-\frac{2(\beta-\alpha)}{3c}\Omega_D^{\frac{3}{2}}
-\frac{\lambda_m}{3},
\end{equation}
\begin{equation}\label{21}
\frac{\dot{\Omega_D}}{H}=(\lambda_m-\lambda_D-1)\Omega_D^2+(1-\lambda_m)\Omega_D+
\frac{2(\beta-\alpha)}{c}\Omega_D^{\frac{3}{2}} (1-\Omega_D).
\end{equation}
For an accelerated universe, $\ddot{a(t)}>0$ (which results in
$\omega<-1/3$), we have
\begin{equation}\label{22}
(1-\lambda_m)+(\lambda_m-\lambda_D-1)\Omega_D-\frac{2(\beta-\alpha)}
{c}\Omega_D^{\frac{3}{2}}<0.
\end{equation}
In the absence of interaction, acceleration is not possible for
$\beta<\alpha$, as was pointed out by \cite{Li} for the case
$\beta=0,\,\alpha=1$. This was the motivation of \cite{Li} to take
the future event horizon (instead of the particle horizon) as the
infrared cutoff. However the above calculation reveals that in the
presence of interaction, inflation may be possible even when the
particle horizon is taken as the infrared cutoff. Combining
(\ref{21}) and (\ref{22}), yields
\begin{equation}\label{23}
\dot{\Omega_D}<\frac{2(\beta-\alpha)}{c}H\Omega_D^{\frac{3}{2}}.
\end{equation}
Hence like the phantom regime, for $\beta<\alpha$ and in an
accelerating universe, $\Omega_D$ is decreasing. Applying the
assumption $\dot{L}>0$, gives an upper bound for $\Omega_D$ which
depends only on $H$
\begin{equation}\label{24}
\dot{\Omega_D}<2H\Omega_D<2H.
\end{equation}
\section{Crossing  $\omega=-1$ in interacting holographic dark
energy model}

At $\omega=-1$, $u=\Omega_D^{\frac{1}{2}}$ satisfies the cubic
equation
\begin{equation}\label{25}
\frac{2(\beta-\alpha)}{c}u^3+(1+\lambda_D-\lambda_m)u^2
+\lambda_m-3=0.
\end{equation}
If $\omega=-1$ is allowed, the above equation must have, at least,
one positive root which is less than one. Based on Descartes rule,
we know that the above equation, at most, has two real positive
roots. So, if $\omega=-1$ is crossed,  two transitions may be
possible, one from quintessence to phantom and the other from
phantom to quintessence phase (by quintessence phase (regime) we
mean $\omega<-1/3$ and $\dot{H}<0$). From (\ref{15}) it is clear
that at $\omega=-1$ we have $\dot{H}=0$. If a transition from
quintessence to phantom phase occurs at time $t_1$, we must have
$\dot{H}(t_1)=0$, and $\dot{H}(t<t_1)<0$ and $\dot{H}(t>t_1)>0$,
therefore $H(t_1)$ must be a local minimum of $H$. In the same
way, $H$ must have a local maximum at $t_2$, where the transition
from phantom to quintessence era occurs. In the neighborhood of
$t_1$, $\omega$ is a decreasing function while in the neighborhood
of $t_2$, $\omega$ is an increasing function of time. So in order
to see that if $\omega=-1$ crossing is permissible, we must also
consider the behavior of the Hubble parameter near the roots of
(\ref{25}).

In the following we assume that $H>0$ and $L>0$. We also assume
that the Hubble parameter is a differentiable function of time
\cite{sad}. Therefore following (\ref{11}) and (\ref{12}), $L$ and
$u(=\Omega_D^{\frac{1}{2}})$ are also differentiable. Let us
consider the Taylor expansion of $H$ at $t=t_i$, where $t_i$ is
defined by $\dot{H}(t_i)=0$ (or $\omega(t_i)=-1$),
\begin{equation}\label{26}
H=h_0+h_1\tau^a+O(\tau^{a+1}),\, a\geq 2,
\end{equation}
where $\tau=t-t_i$, $h_1=\frac{1}{a!}\frac{d^aH}{dt^a}(t_i)$, and
$a$ is the order of the first nonzero derivative of $H$ at
$t=t_i$. If $a$ is an even integer and $h_1>0(<0)$, then $H$ has a
minimum (maximum) at $t_i$ and the transition occurs at $t_i$.
Using (\ref{15}) we obtain
\begin{equation}\label{27}
\omega=-1-\frac{2ah_1}{3h_0^2}\tau^{a-1}+O(\tau^a).
\end{equation}
We consider the following expansion for $u$ at $t=t_i$,
\begin{equation}\label{28}
u=u_0+u_1\tau^b+u_2\tau^{b+1}+O(\tau^{b+2}), \,\, u_1\neq 0, \,\,
b\geq1
\end{equation}
where $b$ is the order of the first nonzero derivative of $u$ at
$t_i$. If the solution (\ref{26}) is permissible, by inserting
(\ref{27}) and (\ref{28}) in (\ref{20}) and by comparing the
powers of $\tau$ in both sides of (\ref{20}), we obtain the
following results:
\begin{equation}\label{29}
(i)-\frac{1+\lambda_D-\lambda_m}{3}u_0^2-\frac{2(\beta-\alpha)}{3c}
u_0^3=-1+\frac{\lambda_m}{3}.
\end{equation}
The roots of this equation specify $u$ at transition time(s). In
order that the transition occurs the above equation must have at
least one real root in the interval $(0,1)$. (ii)For
\begin{equation}\label{30}
\frac{1+\lambda_D-\lambda_m}{-3}-\frac{\beta-\alpha}{c}u_0\neq 0,
\end{equation}
we obtain
\begin{equation}\label{31}
-\frac{ah_1}{3h_0^2}=\left(\frac{1+\lambda_D-\lambda_m}{-3}-
\frac{\beta-\alpha}{c}u_0\right)u_0u_1, \,\, b=a-1.
\end{equation}
In the case
\begin{equation}\label{32}
\frac{1+\lambda_D-\lambda_m}{-3}-\frac{\beta-\alpha}{c}u_0= 0,
\end{equation}
and for $\beta\neq \alpha$, we find
\begin{equation}\label{33}
\frac{2ah_1}{3h_0^2}=\frac{(\beta-\alpha)u_0u_1^2}{c},\,\, a=2b+1.
\end{equation}
In this case $a$ is an odd number and transition does not occur.
If $\beta=\alpha$, $1+\lambda_D-\lambda_m= 0$, and $\omega$ in
(\ref{20}), becomes a constant: $\omega(t)=-1$, and no transition
occurs.

To determine $a$ and the sign of $h_1$, (\ref{21}) may be used :
\begin{equation}\label{34}
2\dot{u}=H\left[\frac{2(\alpha-\beta)}{c}u^4+(\lambda_m-\lambda_D-1)u^3+\frac{2(\beta-\alpha)}{c}u^2
+(1-\lambda_m)u\right].
\end{equation}
By inserting (\ref{28}) and (\ref{26}) into the above equation, we
see that the left hand side begins with $\tau^{b-1}$, while if the
right hand side does not begin with $\tau^0$, it will begin by
$\tau^{(\gamma>b)}$, which is inconsistent with the left hand
side. Thereby $b=1$. Hence based on (\ref{31}) and our previous
discussion after (\ref{26}), if the transition occurs, we must
have $a=2$. Note that for the case (\ref{32}), we obtain $a=3$
which may be corresponding to the inflection point of $H$. For
$b=1$, by equalizing $\tau^0$'s coefficients in both sides of
(\ref{34}), we get
\begin{equation}\label{35}
2u_1=h_0\left[\frac{2(\alpha-\beta)}{c}u_0^4+(\lambda_m-\lambda_D-1)u_0^3
+\frac{2(\beta-\alpha)}{c}u_0^2 +(1-\lambda_m)u_0\right],
\end{equation}
which using (\ref{29}) reduces to
\begin{equation}\label{36}
u_1=h_0u_0\left[\frac{\beta-\alpha}{c}u_0-1\right].
\end{equation}
Now let us determine the sign of $h_1$. Combining (\ref{31}) and
(\ref{36}) results in
\begin{equation}\label{37}
-\frac{ah_1}{3h_0^2}=\left(\frac{1+\lambda_D-\lambda_m}{-3}-\frac{\beta-\alpha}{c}u_0\right)\left(
\frac{\beta-\alpha}{c}u_0-1\right)h_0u_0^2.
\end{equation}
From (\ref{11}), (\ref{12}) and (\ref{13}), by applying the
thermodynamics second law we deduce
\begin{equation}\label{38}
\frac{\beta-\alpha}{c}u_0-1<0,
\end{equation}
which results in $u_1<0$.  $\dot {S}=0$ is ruled out because
$h_1\neq 0$. So we conclude that the sign of $h_1$ must be the
same as the sign of
$(1+\lambda_D-\lambda_m)/(-3)-(\beta-\alpha)u_0/c$. For transition
from quintessence to phantom phase we must have
\begin{equation}\label{39}
\frac{1+\lambda_D-\lambda_m}{-3}-\frac{\beta-\alpha}{c}u_0>0,
\end{equation}
while a transition from phantom to quintessence era requires
\begin{equation}\label{40}
\frac{1+\lambda_D-\lambda_m}{-3}-\frac{\beta-\alpha}{c}u_0<0.
\end{equation}
Note that if, like \cite{abd},  we take $\alpha=0$, $\beta=1$ and
$\lambda_m=\lambda_D$, transition from quintessence to phantom is
not allowed.

It is also instructive to study the behavior of $L$ at $t_i$.
Assuming that $\dot{L}>0$, by inserting the Taylor expansion of
$L$ at $t=t_i$ up to the order $\tau^2$ in (\ref{11}), we obtain
$\dot{L}=[L_0+L_1\tau+O(\tau^2)][h_0+h_1\tau^a+O(\tau^{(a+1)})]+\alpha-\beta$,
where $L_0=L(t_i)$ and $L_1=\dot{L}(t_i)$. Hence
$L_1=h_0L_0+\alpha-\beta$. (\ref{12}) results in $u=c/(LH)$,
therefore
\begin{equation}\label{41}
u_0=\frac{c}{L_0h_0},\,\,u_1=-\frac{c}{L_0}\left(
 1+\frac{\alpha-\beta}{h_0L_0}\right),
\end{equation}
which is consistent with (\ref{36}).

As a summary we have shown that in order that a transition phase
occurs: (i) (\ref{29}) must have at least a positive real root in
the interval (0,1) (ii)  At these roots the Hubble parameter (if
it is differentiable) must have the Taylor expansion (\ref{26}),
with an even integer $a$. In interacting holographic dark energy
model we obtained $a=2$ and verified that quintessence to phantom
phase transition and vice versa occur provided (\ref{39}) and
(\ref{40}) hold respectively.
\section{ Two transitions in interacting holographic dark energy model}
In this part we try to investigate the ability of the system to
return to the quintessence regime from the phantom phase. This may
be interesting because it avoids the big rip singularity which may
be encountered in phantom models.

Let us write (\ref{29}) as
\begin{equation}\label{42}
pu_0^3+qu_0^2+1=0,
\end{equation}
where
\begin{equation}\label{43}
p=\frac{2(\beta-\alpha)}{c(\lambda_m-3)},\,
q=\frac{1+\lambda_D-\lambda_m}{\lambda_m-3}.
\end{equation}
In order to have two transitions, (\ref{42}), must possess two
real positive roots, which we denote by $u_{01}$, $u_{02}$, in the
interval $(0,1)$. $p$ and $q$ are real numbers, hence the third
root must be also real. Therefore the discriminant of (\ref{42}),
 i.e. $-27p^2-4q^3$, must be positive
\begin{equation}\label{44}
\left(\frac{p}{2}\right)^2+\left(\frac{q}{3}\right)^3<0.
\end{equation}
From $\sum_{i\neq j}u_{0i}u_{0j}=0$, we find that the third root,
$u_{03}$, is negative and $|u_{03}|<1$. So using
$0<-u_{03}u_{02}u_{01}=1/p<1$, we deduce $p>1$. Also following
Descartes rule of sign, having two positive roots is only possible
when $p>0$ and $q<0$.

The Sturm sequence constructed from (\ref{42}) is
$\mathcal{S}(x)=\{\mathcal{P}(x),\mathcal{P}_1(x),\mathcal{P}_2(x),\mathcal{P}_3(x)\}$,
where
\begin{eqnarray}\label{45}
\mathcal{P}(x)&=&px^3+qx^2+1 \nonumber \\
\mathcal{P}_1(x)&=&3px^2+2qx \nonumber \\
\mathcal{P}_2(x)&=&\frac{2q^2x}{9p}-1\nonumber \\
\mathcal{P}_3(x)&=&-\frac{9p}{4q^4}(27p^2+4q^3).
\end{eqnarray}
We have $\mathcal{S}(0)=\{1,0,-1,-9p(4q^3+27p^2)/(4q^4)\}$,
$\mathcal{S}(1)=\{p+q+1,3p+2q,-1+2q^2/(9p),-9p(4q^3+27p^2)/(4q^4)\}$.
If we expect to have two real roots in the interval $(0,1)$, using
(\ref{44}) and by applying the Sturm theorem, we find
\begin{eqnarray}\label{46}
&&p+q+1>0 \nonumber \\
&&\frac{p}{2}+\frac{q}{3}>0\nonumber \\
&&\left(\frac{q}{3}\right)^2-\frac{p}{2}>0.
\end{eqnarray}
These conditions and (\ref{44}), may be resumed as
\begin{equation}\label{47}
-q-1<p<2\left( -\frac{q}{3}\right)^{\frac{3}{2}},\,\, q<-3.
\end{equation}

Now let us consider (\ref{39}) and (\ref{40}). A transition from
quintessence to phantom phase may be occurred at $t_1$
($u(t=t_1)=u_{01}$), if $h_1>0$
\begin{equation}\label{48}
\frac{1+\lambda_D-\lambda_m}{-3}-\frac{\beta-\alpha}{c}u_{01}>0.
\end{equation}
For $t>t_1$ the system becomes phantom dominated until $t=t_2$
($u(t=t_2)=u_{02}$), i.e. when the second transition occurs,
provided $h_1<0$ at $t_2$
\begin{equation}\label{49}
\frac{1+\lambda_D-\lambda_m}{-3}-\frac{\beta-\alpha}{c}u_{02}<0.
\end{equation}
From (\ref{14}) we find that $u$ is a decreasing function of time
in phantom dominated era, hence $u_{01}>u_{02}$. So in order that
(\ref{48}) and (\ref{49}) become consistent we must have
\begin{equation}\label{50}
\beta-\alpha<0, \,\, 1+\lambda_D-\lambda_m>0.
\end{equation}

The existence of $u_{01}$ and $u_{02}$ which satisfy (\ref{48})
and (\ref{49}), may be verified as follows: Using the condition
posed on $p$ and $q$, we obtain
\begin{equation}\label{51}
 0<-\frac{2q}{3p}(=-\frac{(1+\lambda_D-\lambda_m)c}{3(\beta-\alpha)})<1.
\end{equation}
The Sturm sequence at $-2q/(3p)$ is
\begin{equation}\label{100}
\mathcal{S}\left(
-2q/(3p)\right)=\{4q^3/(27p^2)+1,0,-4q^3/(27p^2)-1,-9p(4q^3+27p^2)/(4q^4)\}.
\end{equation}
So by invoking the Sturm theorem, one can verify that one of the
positive roots, which based on our discussion in previous
paragraph we take $u_{02}$, is located in $(0, -2q/(3p))$ while
the other, i.e. $u_{01}$ belongs to $(-2q/(3p),1)$. As we have
previously mentioned (see discussion after (\ref{25})), $\omega$
is a decreasing function of time in the neighborhood of $u_{01}$
and an increasing function of time in the vicinity of $u_{02}$,
whence for a differentiable  $\omega$, we must have
$\dot{\omega}=0$ at a point in the phantom regime. It is
interesting to note that for $t_1<t<t_2$, this happens at
$u=-2q/(3p)$, as it can be verified using (\ref{20}). Indeed from
(\ref{20}), it is obvious that $\dot{\omega(t)}=0$ has three
solutions: $u=-2q/(3p)$, $\dot{u}=0$ and $u=0$. In our case,
$\dot{u}=0$ is ruled out by (\ref{23}). The solution $u=-2q/(3p)$,
as we have seen lies in the phantom regime. The solution $u=0$,
must be obtained in the limit $t\to \infty$ in the quintessence
era (by the assumption that the inflation will be continued). To
verify this claim, we note that in an accelerating universe
$\dot{H}+H^2>0$, and for $\alpha>\beta$,
\begin{equation}\label{52}
(HL\dot{)}=(\dot{H}+H^2)L+(\alpha-\beta)H>0.
\end{equation}
Therefore $HL$ is an analytic differentiable increasing function
of time. Thereby $u$, is a decreasing function of time, as we
showed before (see (\ref{23})) via another method. If $u=0$ occurs
at a finite time $\tilde{t}$, then for $t>\tilde{t}$,
$u(t>\tilde{t})<0$ which conflicts with definition of $u$. In
addition $u(\tilde{t})=0$ leads to
$H(\tilde{t})L(\tilde{t})=\infty$ which conflicts with continuity
of $H$ and $L$.

\begin{figure}
\centering\epsfig{file=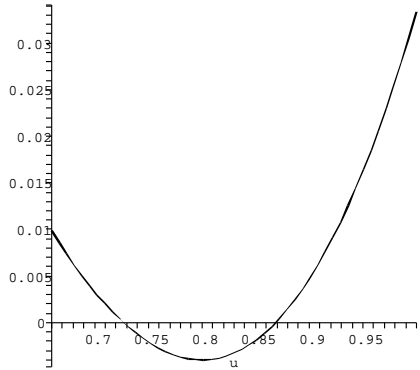,width=6cm} \caption{$\omega+1$ as
a function of $u$, for $\{\beta=0$, $\alpha=1$, $c=1$,
$\lambda_D=3.9$, $\lambda_m=2.5\}$} \label{fig1}
\end{figure}

As a test of our results, we have plotted  $\omega+1$, (\ref{20}),
in terms of $u$ for an interacting holographic dark energy model
with parameters $\{\beta=0$, $\alpha=1$, $c=1$, $\lambda_D=3.9$,
$\lambda_m=2.5\}$ (see fig.(\ref{fig1})). In this example
$\omega+1$ has two zero in the interval $(0,1)$, $u_{01}=0.86$
corresponding to $\Omega_D=0.75$ and $u_{02}=0.72$ corresponding
to $\Omega_D=0.52$. For $u_{02}<u<u_{01}$, the system is in
phantom phase, i.e. $\omega<-1$, and for $u>u_{01}$ and $u<u_{02}$
the system is in quintessence phase. Note that for an accelerating
universe and when $\alpha>\beta$, $u$ is a decreasing function of
time (see(\ref{23}) and our discussion in the previous paragraph)
and the directions of $t$ and $u$-axis are opposite.
$\dot{\omega}=0$ occurs in the phantom regime: $u=0.8$, and at
$u=0$.

\section{Summary}
In this paper we considered the holographic dark energy model with
a general interaction between dark matter and dark energy, see
(\ref{3}). We took the infrared cutoff as a linear combination of
the future an particle horizon see (\ref{9}). We derived an
expression for EOS parameter of the universe, $\omega$, in terms
of the ratio of dark energy density and total energy density of
the flat FRW space-time, $\Omega_D$, see (\ref{20}). Using
(\ref{20}) and the expression obtained for time derivative of
$\Omega_D$ (\ref{21}), and by assumption that the thermodynamics
second law is still valid, we studied the possibility (and
necessary conditions) for quintessence to phantom phase transition
and vice versa with a differentiable Hubble parameter. Using some
theorems about solutions of cubic equation satisfied by
$\Omega_D^{1/2}$ see (\ref{25}), we showed that such transitions
occur provided we appropriately choose the parameters of the
system see(\ref{47}) and (\ref{50}).

\end{document}